\documentclass[12pt]{article}
\usepackage{amssymb}
\usepackage{amsmath}
\usepackage{amsthm}
\usepackage{authblk}
\usepackage{color}
\usepackage{comment}
\usepackage{geometry}		
\usepackage{graphicx}

\DeclareFontFamily{OT1}{pzc}{}
\DeclareFontShape{OT1}{pzc}{m}{it}{<-> s * [1.10] pzcmi7t}{}
\DeclareMathAlphabet{\mathpzc}{OT1}{pzc}{m}{it}

\let\originalleft\left
\let\originalright\right
\renewcommand{\left}{\mathopen{}\mathclose\bgroup\originalleft}
\renewcommand{\right}{\aftergroup\egroup\originalright}

\begin{document}

\newcommand\cN{\mathcal{N}}

\title{Unstable dimension variability, heterodimensional cycles, and blenders in the border-collision normal form.}
\author[$\dagger$]{P.A.~Glendinning}
\author[$\ddagger$]{D.J.W.~Simpson}
\affil[$\dagger$]{Department of Mathematics, University of Manchester, Manchester, UK}
\affil[$\ddagger$]{School of Mathematical and Computational Sciences, Massey University, Palmerston North, New Zealand}

\maketitle



\begin{abstract}

Chaotic attractors commonly contain periodic solutions with unstable manifolds of different dimensions. This allows for a zoo of dynamical phenomena not possible for hyperbolic attractors. The purpose of this Letter is to demonstrate these phenomena in the border-collision normal form. This is a continuous, piecewise-linear family of maps that is physically relevant as it captures the dynamics created in border-collision bifurcations in diverse applications. Since the maps are piecewise-linear they are relatively amenable to an exact analysis and we are able to explicitly identify parameter values for heterodimensional cycles and blenders. For a one-parameter subfamily we identify bifurcations involved in a transition through unstable dimension variability. This is facilitated by being able to compute periodic solutions quickly and accurately, and the piecewise-linear form should provide a useful test-bed for further study.

\end{abstract}

\section{Differing dimensions of instability}
\label{sec:intro}

Chaotic attractors of one-dimensional non-invertible maps and two-dimensional invertible maps have one unstable direction locally.
For higher dimensional maps the dimensions of the unstable manifolds of periodic orbits within a chaotic attractor can differ,
and this can occur also for ODEs.
This phenomenon is known as {\em unstable dimension variability} (UDV).
It is expected to be common for
chaotic attractors in mathematical models with sufficiently
many variables \cite{SaSa18,ViGr00}.
UDV implies  the existence of orbits that spend arbitrarily long times close to an unstable manifold
of one dimension, and arbitrarily long times close to an unstable manifold
of another dimension \cite{KoKa97,DaYo17}.
It follows that finite-time Lyapunov exponents fluctuate about zero as the system evolves \cite{DaGr94,ViBa04}.
It further follows that numerical solutions may differ wildly from actual orbits.
This lack of `shadowing' is problematic for the applicability of a mathematical model \cite{LaGr99,DoLa04}.

One mechanism that implies UDV is the existence of a {\em heterodimensional cycle}
--- a heteroclinic connection between saddle objects
with unstable manifolds of different dimensions.
If an attractor contains a heterodimensional cycle, then it has UDV \cite{BoDi05}.
A given heterodimensional cycle is at least codimension-one, and 
recent advances in numerical methods have led to the identification of
heterodimensional cycles in an ODE model of intracellular calcium dynamics \cite{ZhKr12,HaKr22}.

A consequence of UDV is that
the dimension of a stable or unstable manifold can appear to be greater than that of its local component.
Such manifolds are termed {\em blenders} \cite{BoDi08,BoDi12}.
Blenders form from heteroclinic cycles between invariant sets where at least one of these is chaotic
and the existence of a heterodimensional cycle, and hence UDV, persists as parameters are varied.
In such systems the codimension-one intersection between stable and unstable manifolds is made persistent
if the $m$-dimensional invariant manifold of one of the sets aligns so that it has a projection that
is effectively $(m+1)$-dimensional.
Blenders in some ways generalise the Smale horseshoe to higher dimensions and
provide a way to establish the robustness of properties relating to a lack of hyperbolicity \cite{BoDi05}.
Recently Hittmeyer {\em et.~al.}~\cite{HiKr18,HiKr20} numerically identified blenders in a smooth three-dimensional map.

\section{The border-collision normal form}
\label{sec:bcnf}

Border-collision bifurcations (BCBs) occur when a fixed point of a piecewise-smooth map
collides with a boundary (switching manifold) where the functional form of the map changes.
They have been identified as the onset of chaos and other dynamics
in applications including power electronics \cite{ZhYa10,SiGl21},
mechanical systems with stick-slip friction \cite{DiKo03,SzOs09},
and economics \cite{PuSu06}.
Near a BCB the dynamics
are well-approximated by a piecewise-linear map \cite{Si16}.
The border-collision normal form (BCNF) then results from a change of coordinates \cite{NuYo92,Di03}.
In two dimensions this form is
\begin{equation}
x \mapsto \begin{cases}
A_L x + b, & x_1 \le 0, \\
A_R x + b, & x_1 \ge 0,
\end{cases}
\label{eq:bcnf}
\end{equation}
where
\begin{align}
A_L &= \begin{bmatrix} \tau_L & 1 \\ -\delta_L & 0 \end{bmatrix}, &
A_R &= \begin{bmatrix} \tau_R & 1 \\ -\delta_R & 0 \end{bmatrix}, &
b &= \begin{bmatrix} 1 \\ 0 \end{bmatrix},
\label{eq:ALARb2d}
\end{align}
and $x = (x_1,x_2) \in \mathbb{R}^2$.
Here $\tau_L,\delta_L,\tau_R,\delta_R \in \mathbb{R}$ are parameters;
the BCB parameter, usually denoted $\mu$, has been scaled to $1$.

\begin{figure}
\begin{center}
\includegraphics[width=8.5cm]{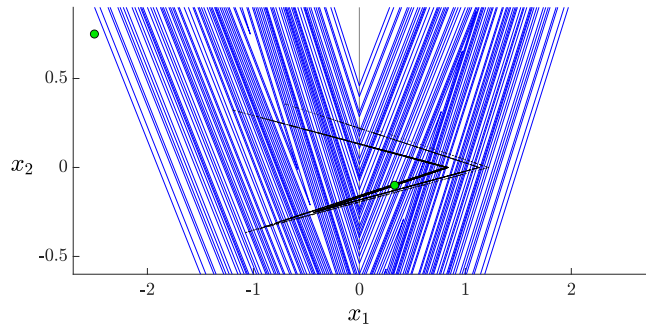}
\end{center}
\caption{
A phase portrait of an invertible instance of the two-dimensional BCNF \eqref{eq:bcnf}--\eqref{eq:ALARb2d};
specifically $(\tau_L,\delta_L,\tau_R,\delta_R) = (1.7,0.3,-1.7,0.3)$.
In black we show $8000$ consecutive iterates of a typical forward orbit with transients removed (this represents the attractor of the map).
The green circles are fixed points; the blue lines show the stable manifold
(grown outwards numerically some amount) of the right-most fixed point.
\label{fig:qqPWLBlender}
}
\end{figure}

The dynamics and bifurcation structure of \eqref{eq:bcnf}--\eqref{eq:ALARb2d}
is incredibly rich \cite{BaGr99,SiMe08b,GlSi22b}.
It exhibits robust chaos \cite{BaYo98} in the sense that
chaotic attractors exist throughout open regions of (four-dimensional) parameter space,
even with $\delta_L \delta_R > 0$ where the map is invertible.
Fig.~\ref{fig:qqPWLBlender} shows a phase portrait of such an attractor.
The attractor contains a saddle fixed point,
and, remarkably, its stable manifold is dense in an open region of phase space.
This was proved in \cite{GhSi22b} via a series of geometric arguments
by bounding the rate at which line segments expand.
A similar result was obtained earlier by Misiurewicz \cite{Mi80}
for the Lozi family (the special case $\tau_L = -\tau_R$ and $\delta_L = \delta_R$).
Despite being one-dimensional, the stable manifold densely fills a two-dimensional region,
so in this sense is a blender.
Hence for two-dimensional, invertible maps,
blenders are possible if the map is $C^0$ but not $C^1$.

\section{Transition through unstable dimension variability}
\label{sec:udv}

The parameter space of the two-dimensional BCNF
has regions where the map has a chaotic attractor
in which (the dense set of) periodic solutions are all saddles,
and other regions where the map has a chaotic attractor
in which periodic solutions are all repellers.
In this section we interpolate between two such regions
and provide numerical evidence for robust UDV.
  
Let $x \in \mathbb{R}^2$ be a period-$n$ point of \eqref{eq:bcnf}--\eqref{eq:ALARb2d}
and suppose its forward orbit does not intersect the switching manifold (as is generically the case).
In a neighbourhood of the orbit the map is differentiable (in fact affine).
Thus it has two stability multipliers,
and if neither of these has modulus $1$, the orbit is hyperbolic.
In this case let $k \in \{0,1,2\}$ denote the number of stability multipliers with modulus greater than $1$.
Then $x$ is asymptotically stable if $k=0$, a saddle if $k=1$, and a repeller if $k=2$.

\begin{figure}[b!]
\begin{center}
\setlength{\unitlength}{1cm}
\begin{picture}(16.53,5.3333)
\put(0,0){\includegraphics[height=5.3333cm]{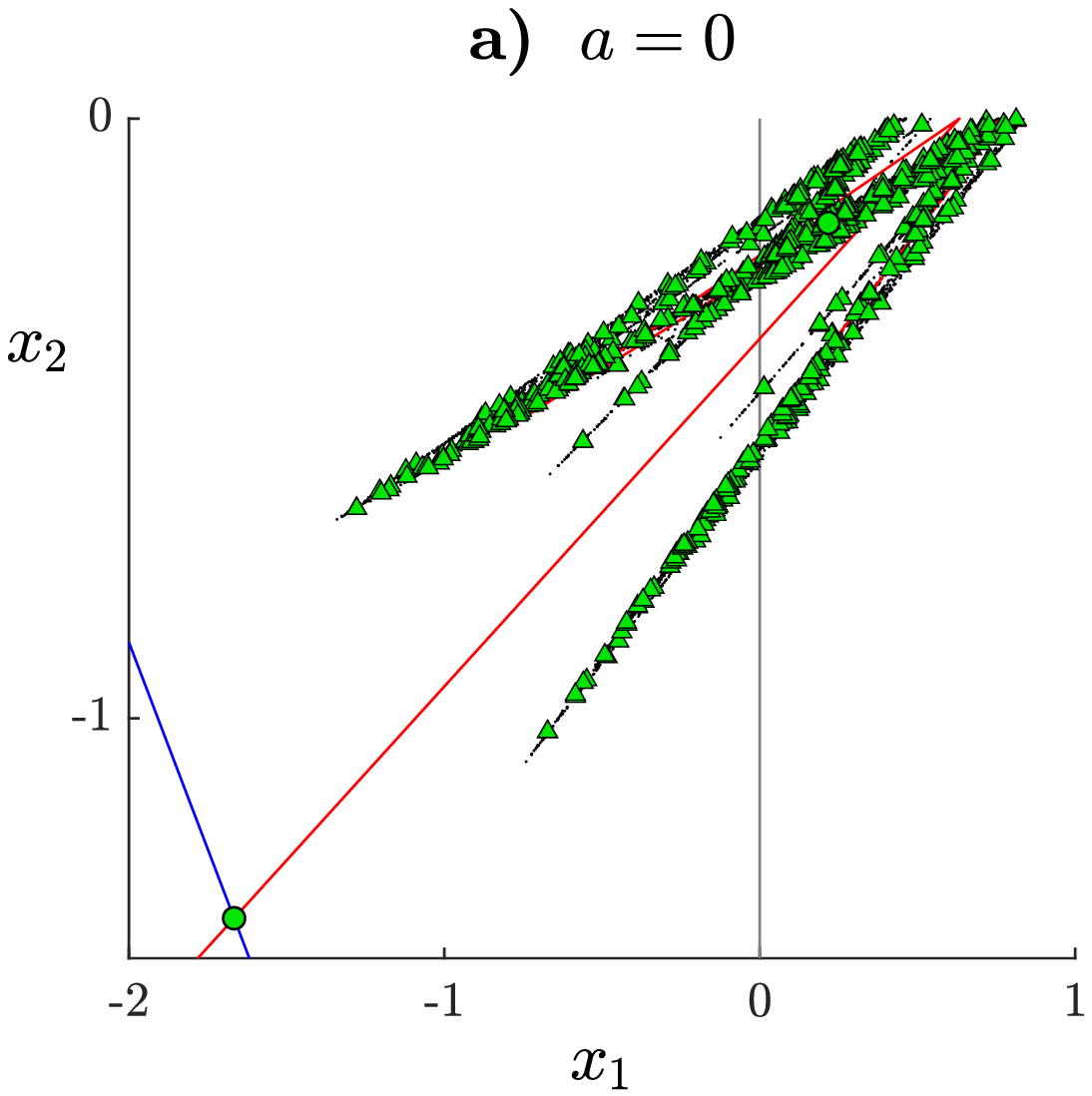}}
\put(5.6,0){\includegraphics[height=5.3333cm]{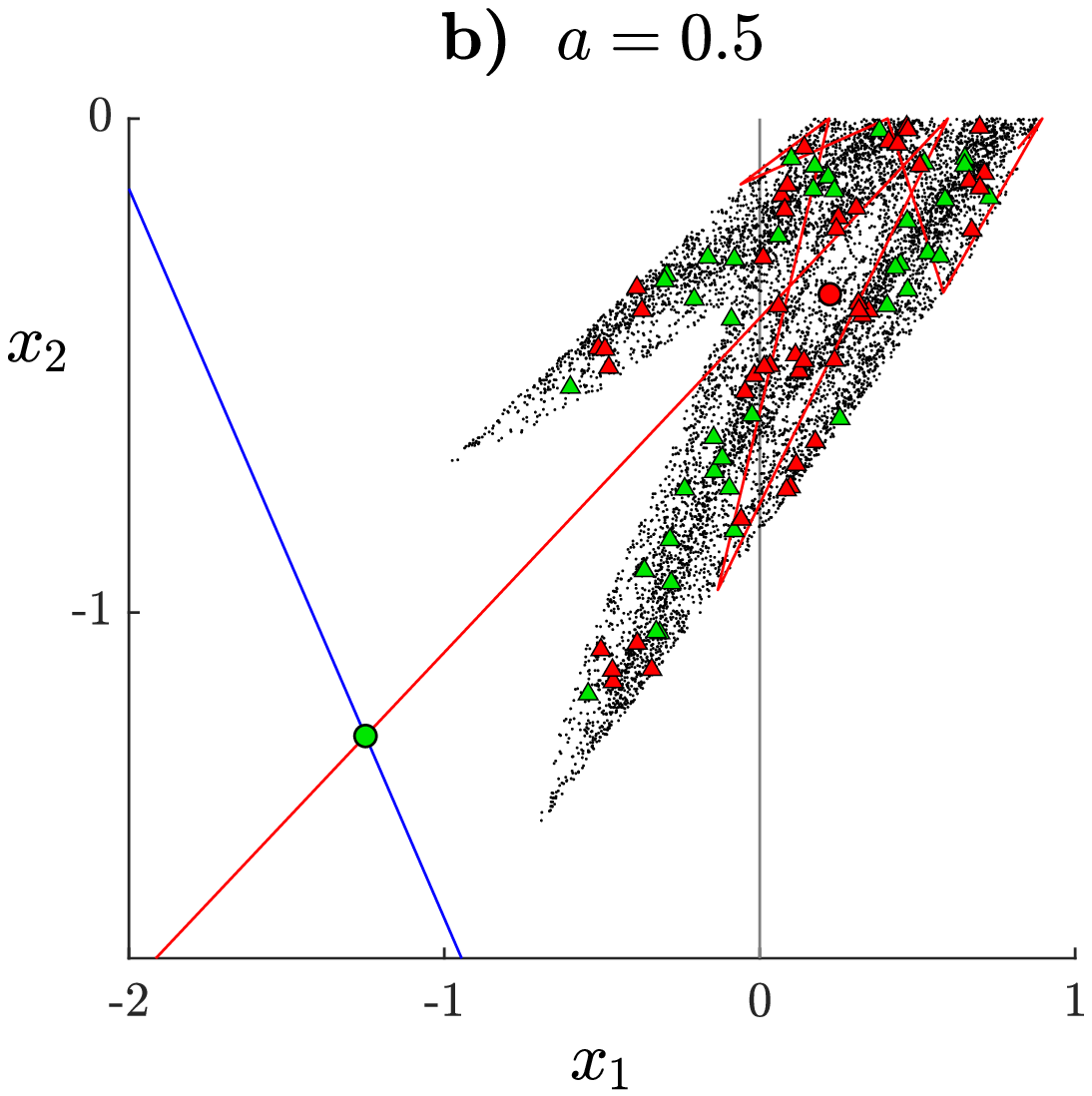}}
\put(11.2,0){\includegraphics[height=5.3333cm]{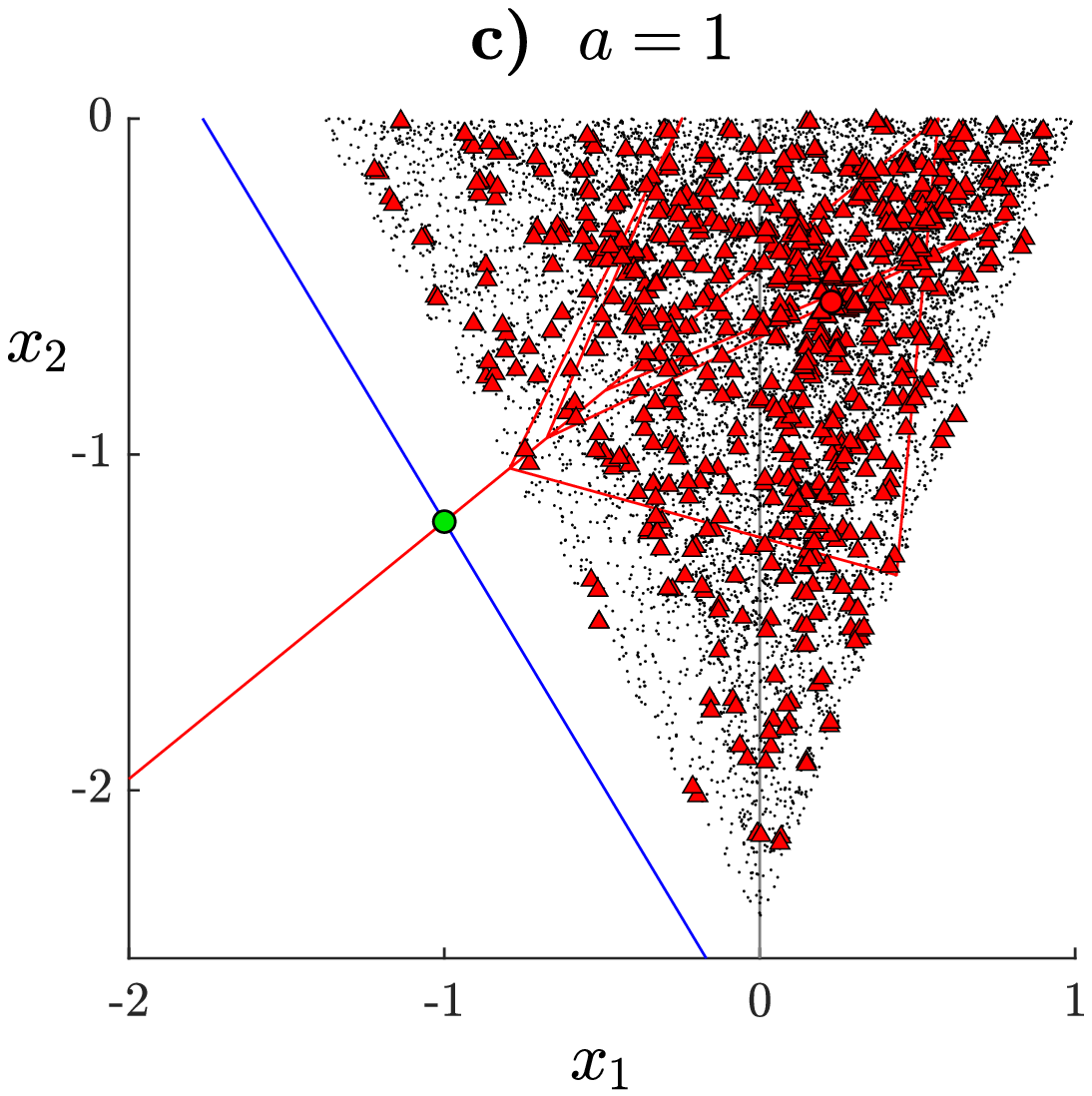}}
\end{picture}
\caption{
Phase portraits of non-invertible instances of the two-dimensional BCNF \eqref{eq:bcnf}--\eqref{eq:ALARb2d}.
The parameter values are given by \eqref{eq:pathDefn} with \eqref{eq:pathEnds}
and three different values of $a$.
The black dots show iterates of a typical forward orbit with transients removed.
Periodic points (up to period $10$) are shown with triangles, except fixed points are shown with circles.
Saddles are coloured green; repellers are coloured red.
The stable (blue) and unstable (red) manifolds of the left-most fixed point
are also shown (grown outwards a small amount).
\label{fig:qqPath}
}
\end{center}
\end{figure}

\begin{figure}
\begin{center}
\includegraphics[width=8.5cm]{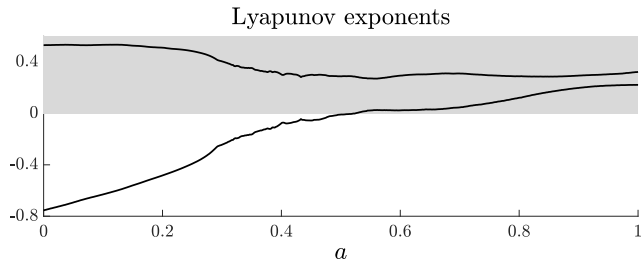}
\end{center}
\caption{
Numerically computed Lyapunov exponents of the attractor of \eqref{eq:bcnf}--\eqref{eq:ALARb2d}
with \eqref{eq:pathDefn} and \eqref{eq:pathEnds}.
\label{fig:lyapMany}
}
\end{figure}

We now explore a one-parameter family of examples.
In \eqref{eq:ALARb2d} we use
\begin{equation}
\begin{aligned}
\tau_L &= (1-a) \tau_{L,0} + a \tau_{L,1} \,, \\
\delta_L &= (1-a) \delta_{L,0} + a \delta_{L,1} \,, \\
\tau_R &= (1-a) \tau_{R,0} + a \tau_{R,1} \,, \\
\delta_R &= (1-a) \delta_{R,0} + a \delta_{R,1} \,,
\label{eq:pathDefn}
\end{aligned}
\end{equation}
with $0 \le a \le 1$ and
\begin{equation}
\begin{aligned}
\tau_{L,0} &= 0.8, & \tau_{L,1} &= 0.8, \\
\delta_{L,0} &= -0.8, & \delta_{L,1} &= -1.2, \\
\tau_{R,0} &= -2.8, & \tau_{R,1} &= -1, \\
\delta_{R,0} &= 0.8, & \delta_{R,1} &= 2.4.
\label{eq:pathEnds}
\end{aligned}
\end{equation}
This one-parameter family has been chosen for three reasons.
First, with $a=0$ all periodic solutions are saddles, Fig.~\ref{fig:qqPath}-a.
This is because with $|\delta_L|, |\delta_R| < 1$ both pieces of \eqref{eq:bcnf} are area-contracting
so repellers are not possible,
while stable periodic solutions are not possible because
an invariant expanding cone can be constructed in tangent space \cite{GlSi21}.
Second, with $a=1$ all periodic solutions, except the left-most fixed point, appear to be repellers, Fig.~\ref{fig:qqPath}-c.
This has been proved for nearby parameter combinations
where there exists a simple Markov partition \cite{GlWo11,Gl16e}.
Third, the map appears to have a unique attractor for all $0 \le a \le 1$.
The two Lyapunov exponents of the attractor are shown in Fig.~\ref{fig:lyapMany}.
These were computed numerically using the standard QR-factorisation method \cite{EcRu85,VoUd97}.

\begin{figure}
\begin{center}
\setlength{\unitlength}{1cm}
\begin{picture}(8.5,4)
\put(0,0){\includegraphics[height=4cm]{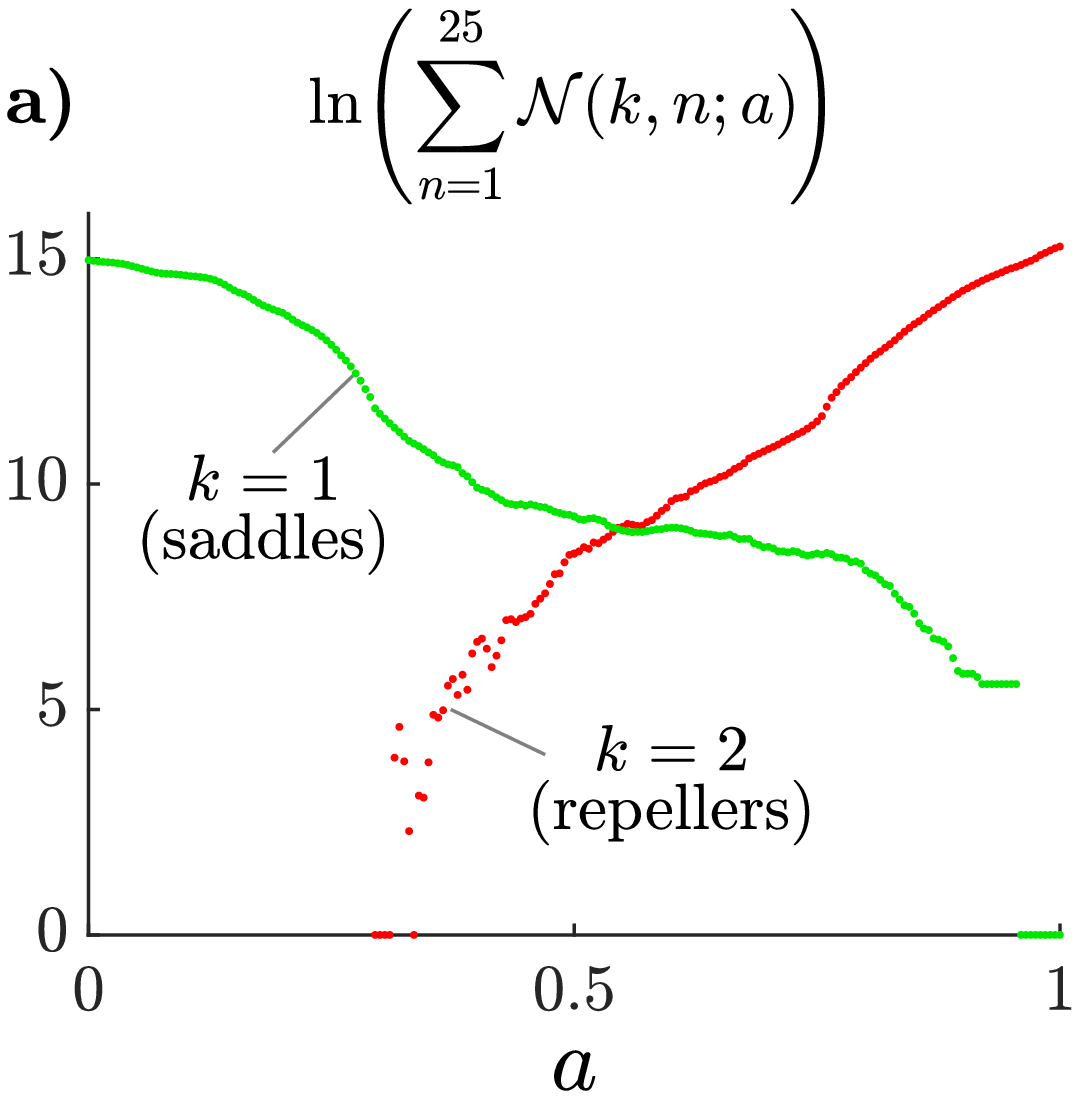}}
\put(4.5,0){\includegraphics[height=4cm]{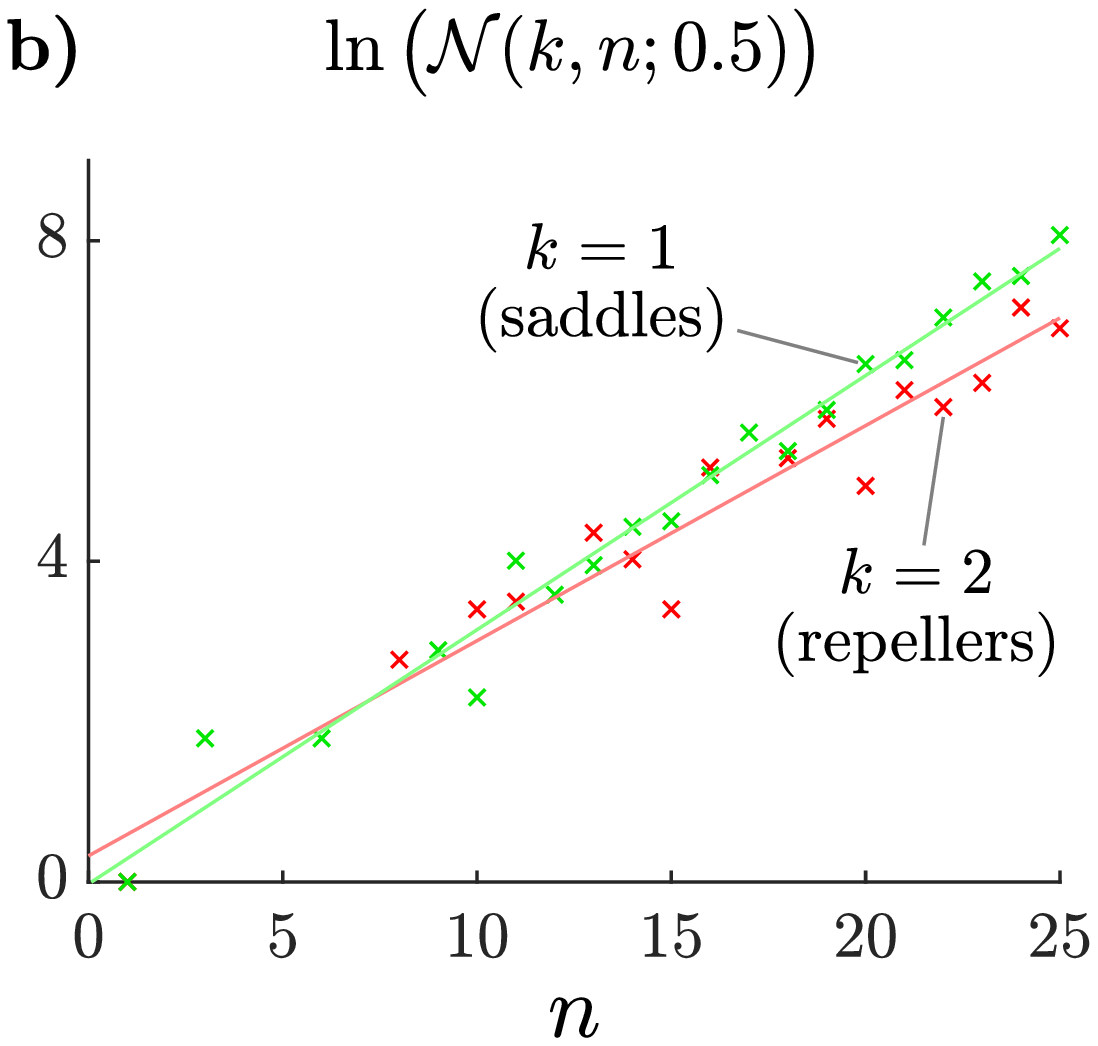}}
\end{picture}
\caption{
Plots involving $\cN(k,n;a)$: the number of period-$n$ points (saddles for $k=1$; repellers for $k=2$)
of \eqref{eq:bcnf}--\eqref{eq:ALARb2d} with \eqref{eq:pathDefn} and \eqref{eq:pathEnds}.
Panel (b) includes lines of best fit.
\label{fig:countPerSolns}
}
\end{center}
\end{figure}

For any $0 \le a \le 1$,
let $\cN(k,n;a)$ denote the number of period-$n$ points that have $k$ stability multipliers with modulus greater than $1$.
The sum of these numbers up to $n=25$ is plotted in Fig.~\ref{fig:countPerSolns}-a.
This figure was computed by brute-force.
We used Duval's algorithm \cite{Du88} to generate all sequences of $L$'s and $R$'s of length $n \le 25$.
Interpreting these as applications of \eqref{eq:bcnf} on the left or on the right respectively,
the (generically) unique point that has period $n$ in the order specified by each sequence was 
identified for each value of $a$ \cite{Si16}. 
We then checked by iterating the map whether the order of the images of the point matched 
the order of the specified sequence (an {\em admissibility} condition).
For those admissible sequences
the stability multipliers were evaluated to determine whether the periodic solution is a saddle or a repeller.

For an intermediate range of values of $a$,
the attractor contains both saddles and repellers, thus exhibits UDV.
The point of cross-over,
where saddles and repellers exist in the same proportion,
is close to $a = 0.5$ and matches well to where the lower Lyapunov exponent becomes positive.
Fig.~\ref{fig:qqPath}-b shows a phase portrait with $a = 0.5$;
Fig.~\ref{fig:countPerSolns}-b shows that here the number of saddles
and repellers appears to increase exponentially with $n$.
This suggests that saddles and repellers are both dense in the attractor.

Saddles and repellers can arise in different ways.
As the value of $a$ is decreased from $1$,
saddles appear to be first created in a BCB
of period-five solutions at $a \approx 0.9592$.
Here infinitely many saddle periodic points are created because the BCB
also creates robust heteroclinic connections between two saddle period-five solutions.
This explains the large discontinuity in the number of saddle points in Fig.~\ref{fig:countPerSolns}-a.
In contrast, as the value of $a$ is increased from $0$,
repellers are created and destroyed in many bifurcations.
For example a saddle period-nine solution (with only one point in $x_1 < 0$)
becomes repelling at $a \approx 0.3278$ when one of its stability multipliers decreases through $-1$,
then is destroyed in a BCB at $a \approx 0.3321$.

Since the maps are non-invertible, repelling sets may have preimages,
i.e.~they may have zero dimensional stable manifolds leading to phenomena such as snap-back repellers \cite{Gl10}.
We conjecture that the saddle chaotic set associated with the period-five heteroclinic cycle
at $a \approx 0.9592$ provides the larger set required to give robust intersections between its one-dimensional unstable manifold
and the zero-dimensional stable manifolds of the repellers as $a$ decreases further,
thus creating the conditions for blenders to exist.
The transition at lower values of $a$ to the appearance of repellers is less clear.
It may be that `enough' repellers need to be created before a persistent connection between
their zero-dimensional stable manifolds and the unstable manifolds of the saddles can be created.

\section{An explicit heterodimensional cycle}
\label{sec:hd}

\begin{figure}[b!]
\begin{center}
\includegraphics[width=8.5cm]{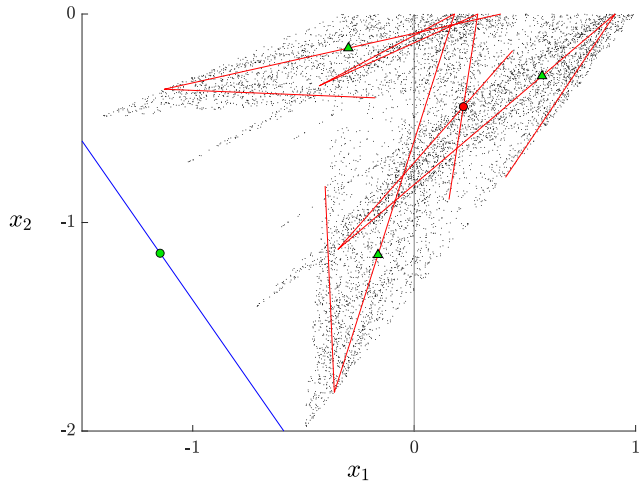}
\end{center}
\caption{
A phase portrait of a non-invertible instance of the two-dimensional BCNF,
\eqref{eq:bcnf}--\eqref{eq:ALARb2d} with \eqref{eq:param_hdcycle_3_to_1}.
The black dots show iterates of a typical forward orbit with transients removed.
The blue line is the initial part of the stable manifold of the left-most fixed point (green circle).
The red lines show part of the unstable manifold of a period-three solution (green triangles).
The value of $\tau_L$ has been chosen so that this manifold intersects the right-most fixed point (red circle).
\label{fig:hdcycle_3_to_1}
}
\end{figure}

We now provide a simple example of a heterodimensional cycle.
Fig.~\ref{fig:hdcycle_3_to_1} shows a phase portrait of \eqref{eq:bcnf} with
\begin{align}
\tau_L &\approx 0.8716, &
\delta_L &= -1, &
\tau_R &= -1.5, &
\delta_R &= 2,
\label{eq:param_hdcycle_3_to_1}
\end{align}
where the exact value of $\tau_L$ will be clarified in a moment.
With these values the right-most fixed point (red circle), call it $x^R$, is repelling.
There also exists a saddle period-three solution (green triangles).
The value of $\tau_L$ has been chosen so that the unstable manifold of the period-three solution intersects $x^R$.
By using computer algebra to analytically find where a certain fourth preimage of $x^R$
lies on the initial linear part of the unstable manifold,
we found that $\tau_L$ is a root of
\begin{equation}
108 \tau_L^6 + 495 \tau_L^5 + 258 \tau_L^4 + 1184 \tau_L^3 - 5800 \tau_L^2 - 4907 \tau_L + 7454.
\label{eq:polynomial_hdcycle_3_to_1}
\end{equation}

The (two-dimensional) unstable manifold of $x^R$
appears to intersect the stable manifold of the period-three solution (as one would expect),
thus these orbits have a heteroclinic connection.
This is a heterodimensional cycle because
$x^R$ and the period-three solution have unstable manifolds of different dimensions.
This cycle is codimension-one because their dimensions differ by one;
indeed the cycle was obtained by carefully adjusting the value of one parameter (namely $\tau_L$).

Fig.~\ref{fig:hdcycle_3_to_1} is the simplest example of a heterodimensional cycle 
that we have found for \eqref{eq:bcnf}--\eqref{eq:ALARb2d} where the cycle is contained in an attractor.
This suggests that, as in Fig.~\ref{fig:qqPath}-b, the attractor exhibits UDV.

\section{Unstable dimension variability in invertible maps}
\label{sec:3d}

For an invertible map to have a heterodimensional cycle, the map needs to be at least three dimensional.
This can also be demonstrated with the BCNF.
In three dimensions the BCNF is \eqref{eq:bcnf} with
\begin{align}
A_L &= \begin{bmatrix} \tau_L & 1 & 0 \\ -\sigma_L & 0 & 1 \\ \delta_L & 0 & 0 \end{bmatrix}, &
A_R &= \begin{bmatrix} \tau_R & 1 & 0 \\ -\sigma_R & 0 & 1 \\ \delta_R & 0 & 0 \end{bmatrix}, &
b &= \begin{bmatrix} 1 \\ 0 \\ 0 \end{bmatrix},
\label{eq:ALARb3d}
\end{align}
and $x = (x_1,x_2,x_3) \in \mathbb{R}^3$,
and has been studied previously for instance in \cite{DeDu11,Si17d}.

\begin{figure}[b!]
\begin{center}
\includegraphics[width=13cm]{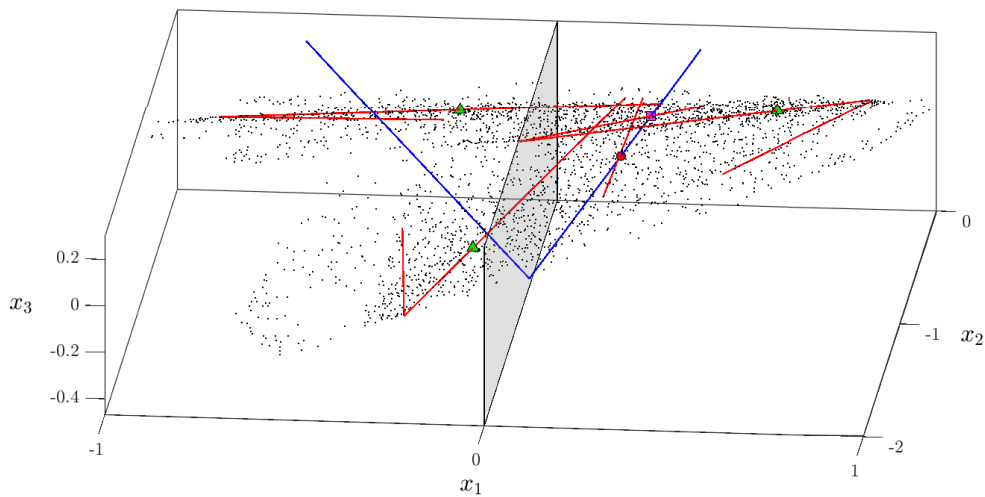}
\end{center}
\caption{
A phase portrait of an invertible instance of the three-dimensional BCNF,
\eqref{eq:bcnf} with \eqref{eq:ALARb3d} and \eqref{eq:param_hdcycle3d_3_to_1}.
The black dots show iterates of a typical forward orbit with transients removed.
The one-dimensional stable manifold of a fixed point (red circle)
approximately intersects the one-dimensional unstable manifold of a period-three solution (green triangles).
One point of intersection is indicated with a pink square.
\label{fig:hdcycle3d_3_to_1a}
}
\end{figure}

Fig.~\ref{fig:hdcycle3d_3_to_1a} shows a phase portrait using
\begin{equation}
\begin{aligned}
\tau_L &= 0.7228540306, &
\sigma_L &= -1, &
\delta_L &= -0.2, \\
\tau_R &= -1.5, &
\sigma_R &= 2, &
\delta_R &= -0.2.
\end{aligned}
\label{eq:param_hdcycle3d_3_to_1}
\end{equation}
These values were obtained by adding a dimension to the example of Fig.~\ref{fig:hdcycle_3_to_1},
varying $\delta_L$ and $\delta_R$ from $0$ to create fully three-dimensional dynamics,
and lastly adjusting the value of $\tau_L$ (to 10 decimal places)
so that the one-dimensional unstable manifold of the period-three solution
approximately intersects the one-dimensional stable manifold of $x^R$.
The pink square in Fig.~\ref{fig:hdcycle3d_3_to_1a} shows this approximate point of intersection.
Since the invariant manifolds appear to be embedded in an attractor,
the other invariant manifolds presumably intersect forming a heterodimensional cycle.
Hence this attractor too has UDV.

\begin{figure}[b!]
\begin{center}
\includegraphics[width=8.5cm]{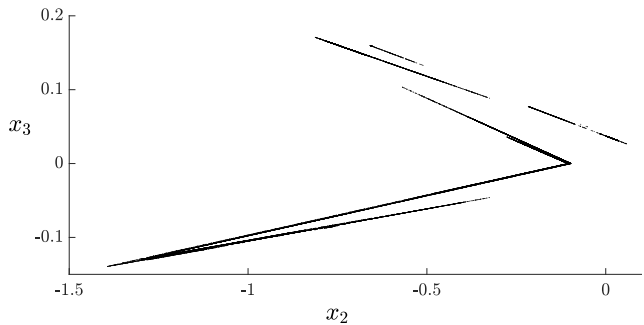}
\end{center}
\caption{
Intersections of the one-dimensional unstable manifold of
the period-three solution of Fig.~\ref{fig:hdcycle3d_3_to_1a} with $x_1 = 0$.
This was computed by growing the manifold much further than that shown in Fig.~\ref{fig:hdcycle3d_3_to_1a}.
\label{fig:hdcycle3d_3_to_1b}
}
\end{figure}

Numerically we grown the unstable manifold of the period-three solution
outwards much further than that shown in Fig.~\ref{fig:hdcycle3d_3_to_1a}.
Fig.~\ref{fig:hdcycle3d_3_to_1b} shows the intersections of this manifold with the switching manifold $x_1 = 0$.
The intersections have a quasi-one-dimensional structure
suggesting that the unstable manifold is essentially two-dimensional, so a blender.

\section{Discussion}
\label{sec:conc}

The existence of UDV due to blenders has been established numerically
in three-dimensional generalisations of the H\'enon map \cite{SaSa18,SaTa21}.
We have considered a related piecewise-linear family and
by interpolating between parameters with only saddles and parameters with only repellers
we have provided strong numerical evidence for the existence of UDV in the two-dimensional non-invertible BCNF
and the three-dimensional invertible BCNF.
Since the BCNF describes BCBs in general piecewise-smooth systems
the existence of UDV in these examples shows that UDV has broader significance
within the study of piecewise-smooth dynamical systems and their applications.

We have also identified a possible mechanism for the onset of UDV through the creation of
saddle chaotic sets (as the parameter $a$ in Fig.~\ref{fig:countPerSolns} decreases) and snap-back repellers (as $a$ increases).
These may provide stable and unstable manifolds of the appropriate dimensions to create blenders
with the infinite families of periodic orbits (either repellers or saddles) that already exist.
This conjectural connection merits further study, with comparisons to the bifurcation theory approaches of \cite{AlSa06,BaSo00}.

\section*{Acknowledgements}
\setcounter{equation}{0}

The authors were supported by Marsden Fund contract MAU1809, managed by Royal Society Te Ap\={a}rangi.


\begin{thebibliography}{10}

\bibitem{AlSa06}
K.T. Alligood, E.~Sander, and J.A. Yorke.
\newblock Crossing bifurcations and unstable dimension variability.
\newblock {\em Phys. Rev. Lett.}, 96:244103, 2006.

\bibitem{BaGr99}
S.~Banerjee and C.~Grebogi.
\newblock Border collision bifurcations in two-dimensional piecewise smooth
  maps.
\newblock {\em Phys. Rev. E}, 59(4):4052--4061, 1999.

\bibitem{BaYo98}
S.~Banerjee, J.A. Yorke, and C.~Grebogi.
\newblock Robust chaos.
\newblock {\em Phys. Rev. Lett.}, 80(14):3049--3052, 1998.

\bibitem{BaSo00}
E.~Barreto and P.~So.
\newblock Mechanisms for the development of unstable dimension variability and
  the breakdown of shadowing in coupled chaotic systems.
\newblock {\em Phys. Rev. Lett.}, 85(12):2490--2493, 2000.

\bibitem{BoDi08}
C.~Bonatti and L.J. D\'{\i}az.
\newblock Robust heterodimensional cycles and {$C^1$}-generic dynamics.
\newblock {\em J. Inst. Math. Jussieu}, 7:469--525, 2008.

\bibitem{BoDi12}
C.~Bonatti, L.J. D\'{\i}az, and S.~Kiriki.
\newblock Stabilization of heterodimensional cycles.
\newblock {\em Nonlinearity}, 25:931--960, 2012.

\bibitem{BoDi05}
C.~Bonatti, L.J. D\'{\i}az, and M.~Viana.
\newblock {\em Dynamics Beyond Uniform Hyperbolicity.}
\newblock Springer, New York, 2005.

\bibitem{DaYo17}
S.~Das and J.A. Yorke.
\newblock Multichaos from quasiperiodicity.
\newblock {\em SIAM J. Appl. Dyn. Syst.}, 16(4), 2017.

\bibitem{DaGr94}
S.~Dawson, C.~Grebogi, T.~Sauer, and J.A. Yorke.
\newblock Obstructions to shadowing when a {L}yapunov exponent fluctuates about
  zero.
\newblock {\em Phys. Rev. Lett.}, 73(14):1927--1930, 1994.

\bibitem{DeDu11}
S.~De, P.S. Dutta, S.~Banerjee, and A.R. Roy.
\newblock Local and global bifurcations in three-dimensional, continuous,
  piecewise-smooth maps.
\newblock {\em Int. J. Bifurcation Chaos}, 21(6):1617--1636, 2011.

\bibitem{Di03}
M.~di~Bernardo.
\newblock Normal forms of border collision in high dimensional non-smooth maps.
\newblock In {\em Proceedings IEEE ISCAS, Bangkok, Thailand}, volume~3, pages
  76--79, 2003.

\bibitem{DiKo03}
M.~di~Bernardo, P.~Kowalczyk, and A.~Nordmark.
\newblock Sliding bifurcations: {A} novel mechanism for the sudden onset of
  chaos in dry friction oscillators.
\newblock {\em Int. J. Bifurcation Chaos}, 13(10):2935--2948, 2003.

\bibitem{DoLa04}
Y.~Do and Y.-C. Lai.
\newblock Statistics of shadowing time in nonhyperbolic chaotic systems with
  unstable dimension variability.
\newblock {\em Phys. Rev. E}, 69:016213, 2004.

\bibitem{Du88}
J.-P. Duval.
\newblock G\'{e}n\'{e}ration d'une section des classes de conjugaison et arbre
  des mots de {L}yndon de longueur born\'{e}e.
\newblock {\em Theoret. Comput. Sci.}, 60:255--283, 1988.
\newblock In French.

\bibitem{EcRu85}
J.-P. Eckmann and D.~Ruelle.
\newblock Ergodic theory of chaos and strange attractors.
\newblock {\em Rev. Mod. Phys.}, 57(3):617--656, 1985.

\bibitem{GhSi22b}
I.~Ghosh and D.J.W. Simpson.
\newblock Robust {D}evaney chaos in the two-dimensional border-collision normal
  form.
\newblock {\em Chaos}, 32:043120, 2022.

\bibitem{Gl10}
P.~Glendinning.
\newblock Bifurcations of snap-back repellers with application to
  border-collision bifurcations.
\newblock {\em Int. J. Bifurcation Chaos}, 20:479--489, 2010.

\bibitem{Gl16e}
P.~Glendinning.
\newblock Bifurcation from stable fixed point to {2D} attractor in the border
  collision normal form.
\newblock {\em IMA J. Appl. Math.}, 81(4):699--710, 2016.

\bibitem{GlSi22b}
P.~Glendinning and D.J.W. Simpson.
\newblock Chaos in the border-collision normal form: {A} computer-assisted
  proof using induced maps and invariant expanding cones.
\newblock {\em Appl. Math. Comput.}, 434:127357, 2022.

\bibitem{GlWo11}
P.~Glendinning and C.H. Wong.
\newblock Two dimensional attractors in the border collision normal form.
\newblock {\em Nonlinearity}, 24:995--1010, 2011.

\bibitem{GlSi21}
P.A. Glendinning and D.J.W. Simpson.
\newblock A constructive approach to robust chaos using invariant manifolds and
  expanding cones.
\newblock {\em Discrete Contin. Dyn. Syst.}, 41(7):3367--3387, 2021.

\bibitem{HaKr22}
A.~Hammerlindl, B.~Krauskopf, G.~Mason, and H.M. Osinga.
\newblock Determining the global manifold structure of a continuous-time
  heterodimensional cycle.
\newblock To appear: {\em J. Comput. Dyn.}, 2022.

\bibitem{HiKr18}
S.~Hittmeyer, B.~Krauskopf, H.M. Osinga, and K.~Shinohara.
\newblock Existence of blenders in a {H}\'{e}non-like family: geometric
  insights from invariant manifold computations.
\newblock {\em Nonlinearity}, 31(10):R239--R267, 2018.

\bibitem{HiKr20}
S.~Hittmeyer, B.~Krauskopf, H.M. Osinga, and K.~Shinohara.
\newblock How to identify a hyperbolic set as a blender.
\newblock {\em Discrete Cont. Dyn. Syst.}, 40(12):6815--6836, 2020.

\bibitem{KoKa97}
E.J. Kostelich, I.~Kan, C.~Grebogi, E.~Ott, and J.A. Yorke.
\newblock Unstable dimension variability: {A} source of nonhyperbolicity in
  chaotic systems.
\newblock {\em Phys. D}, 109:81--90, 1997.

\bibitem{LaGr99}
Y.-C. Lai and C.~Grebogi.
\newblock Modeling of coupled chaotic oscillators.
\newblock {\em Phys. Rev. Lett.}, 82(24):4803--4806, 1999.

\bibitem{Mi80}
M.~Misiurewicz.
\newblock Strange attractors for the {L}ozi mappings.
\newblock In R.G. Helleman, editor, {\em Nonlinear dynamics, Annals of the New
  York Academy of Sciences}, pages 348--358, New York, 1980. Wiley.

\bibitem{NuYo92}
H.E. Nusse and J.A. Yorke.
\newblock Border-collision bifurcations including ``period two to period
  three'' for piecewise smooth systems.
\newblock {\em Phys. D}, 57:39--57, 1992.

\bibitem{PuSu06}
T.~Puu and I.~Sushko, editors.
\newblock {\em Business Cycle Dynamics: Models and Tools.}
\newblock Springer-Verlag, New York, 2006.

\bibitem{SaSa18}
Y.~Saiki, M.A.F. Sanju\'an, and J.A. Yorke.
\newblock Low-dimensional paradigms for high-dimensional hetero-chaos.
\newblock {\em Chaos}, 28(10):103110, 2018.

\bibitem{SaTa21}
Y.~Saiki, H.~Takahasi, and J.A. Yorke.
\newblock Piecewise-linear maps with heterogeneous chaos.
\newblock {\em Nonlinearity}, 34:5744--5761, 2021.

\bibitem{Si16}
D.J.W. Simpson.
\newblock Border-collision bifurcations in $\mathbb{R}^n$.
\newblock {\em SIAM Rev.}, 58(2):177--226, 2016.

\bibitem{Si17d}
D.J.W. Simpson.
\newblock Grazing-sliding bifurcations creating infinitely many attractors.
\newblock {\em Int. J. Bifurcation Chaos}, 27(12):1730042, 2017.

\bibitem{SiGl21}
D.J.W. Simpson and P.A. Glendinning.
\newblock Inclusion of higher-order terms in the border-collision normal form:
  persistence of chaos and applications to power converters.
\newblock Submitted to: {\em Phys. D}. \texttt{arXiv:2111.12222}, 2021.

\bibitem{SiMe08b}
D.J.W. Simpson and J.D. Meiss.
\newblock Neimark-{S}acker bifurcations in planar, piecewise-smooth, continuous
  maps.
\newblock {\em SIAM J. Appl. Dyn. Sys.}, 7(3):795--824, 2008.

\bibitem{SzOs09}
R.~Szalai and H.M. Osinga.
\newblock Arnol'd tongues arising from a grazing-sliding bifurcation.
\newblock {\em SIAM J. Appl. Dyn. Sys.}, 8(4):1434--1461, 2009.

\bibitem{ViBa04}
R.L. Viana, J.R.R. Barbosa, and C.~Grebogi.
\newblock Unstable dimension variability and codimension-one bifurcations of
  two-dimensional maps.
\newblock {\em Phys. Lett. A}, 321:244--251, 2004.

\bibitem{ViGr00}
R.L. Viana and C.~Grebogi.
\newblock Unstable dimension variability and synchronization of chaotic
  systems.
\newblock {\em Phys. Rev. E}, 62(1):462--468, 2000.

\bibitem{VoUd97}
H.F. von Bremen, F.E. Udwadia, and W.~Proskurowski.
\newblock An efficient {QR} based method for the computation of {L}yapunov
  exponents.
\newblock {\em Phys. D}, 101:1--16, 1997.

\bibitem{ZhKr12}
W.~Zhang, B.~Krauskopf, and V.~Kirk.
\newblock How to find a codimension-one heteroclinic cycle between two periodic
  orbits.
\newblock {\em Discrete Contin. Dyn. Syst.}, 32(8):2825--2851, 2012.

\bibitem{ZhYa10}
Z.T. Zhusubaliyev, O.O. Yanochkina, E.~Mosekilde, and S.~Banerjee.
\newblock Two-mode dynamics in pulse-modulated control systems.
\newblock {\em Annual Rev. Control}, 34:62--70, 2010.

\end{thebibliography}

\end{document}